\documentclass[twocolumn,aps,showpacs]{revtex4}
\usepackage{graphicx}
\usepackage{dcolumn}
\usepackage{bm}
\usepackage{epsf}
\usepackage{epsfig}
\usepackage{amsmath,amsfonts,amsthm, amssymb,latexsym}
\usepackage{enumerate}

\newcommand{\beq}{\begin{equation}}
\newcommand{\eeq}{\end{equation}}
\newcommand{\bcn}{\begin{center}}
\newcommand{\ecn}{\end{center}}

\newcommand{\nmass}{m_{\text{N}}}

\newcommand{\lsim}{\lower0.5ex\hbox{$\; \buildrel < \over \sim \;$}}

\newcommand{\msun}{M_{\odot}}

\begin{document}

\title{Thermal Evolution of Neutron Stars in Two Dimensions}

\date{\today}

\author{Rodrigo Negreiros} \email{negreiros@fias.uni-frankfurt.de}

\author{Stefan Schramm} 
\affiliation{FIAS, Goethe University, Ruth Moufang Str.\ 1,
        60438 Frankfurt, Germany}

\author{Fridolin Weber} 
\affiliation{Department of Physics, San Diego State University, 5500
  Campanile Drive, San Diego, California 92182, USA}

\begin{abstract}
  There are many factors that contribute to the breaking of the
  spherical symmetry of a neutron star. Most notably is rotation,
  magnetic fields, and/or accretion of matter from companion
  stars. All these phenomena influence the macroscopic structures of
  neutron stars, but also impact their microscopic compositions.
  The purpose of this paper is to investigate the cooling of
  rotationally deformed, two-dimensional (2D) neutron stars in the
  framework of general relativity theory, with the ultimate goal of
  better understand the impact of 2D effects on the thermal evolution
  of such objects. The equations that govern the thermal evolution of
  rotating neutron stars are presented in this paper. The cooling of
  neutron stars with different frequencies is computed
  self-consistently by combining a fully general relativistic 2D
  rotation code with a general relativistic 2D cooling code. We show
  that rotation can significantly influence the thermal evolution of
  rotating neutron stars. Among the major new aspects are the
  appearances of hot spots on the poles, and an increase of the
  thermal coupling times between the core and the crust of rotating
  neutron stars. We show that this increase is independent of the
  microscopic properties of the stellar core, but depends only on the
  frequency of the star.
\end{abstract}


\maketitle

\section{Introduction}
The cooling of neutron stars has been used by many authors
\cite{Tsuruta1965,Maxwell1979,Schaab1996,Page2004,Page2006,Page2009,Blaschke2000,Grigorian2005,
  Blaschke2006,Page2011,Yakovlev2011,Voskresensky1997,Shternin2011,Shovkovy2002,Prakash1992,%
  Lattimer1994,Gusakov2005,Horvath1991,Weber2007b,Negreiros2010,Niebergal,Alford2005a}
as a way of probing the internal composition of these objects. Such
studies rely on the fact that the physical quantities relevant for the
cooling (specific heat, thermal conductivity, and neutrino emissions)
strongly depend on the microscopic composition, so that different
models lead to different thermal evolutions. For such studies, the
predicted thermal evolution is compared to the observed data, with the
ultimate goal of constraining the microscopic properties of neutron
stars \cite{Page2004,Page2006}.  An example of this is the recent
analysis of the observed thermal behavior of the compact star in
Cassiopeia A (Cas A) \cite{Page2011,Yakovlev2011,Heinke2010}. In these
studies the authors link the observed thermal behavior of this neutron
star to the onset of superfluidity in the stellar core. Possible
alternative explanations have been suggested in
\cite{Blaschke2011,Negreiros2011}, where the observed data was
explained in terms of the nuclear medium cooling scenario
\cite{Blaschke2011}, and the a late onset of the Direct Urca process,
triggered by the gravitational compression that accompanies a
spinning-down neutron star \cite{Negreiros2011}.

In the standard approach, studies of the thermal evolution of neutron
stars are performed by assuming that these stars are spherically
symmetric. This assumption renders the thermal evolution calculations
inherently one-dimensional (1D), which greatly simplifies the
numerical treatment. However, as pointed out by us in
\cite{Negreiros2011}, rotation may play a very important role for the
cooling of neutron stars. In \cite{Negreiros2011} the effects of
rotation on the microscopic composition, and its consequences for
cooling, were studied. In the work presented here we investigate the
macroscopic aspects of rotation on the cooling of neutron stars in the
framework of a fully self-consistent two-dimensional (2D) treatment.

We shall consider the thermal evolution of rigidly rotating neutron
stars. First we calculate the rotational structures of such stars,
which is considerably more complicated than for spherically symmetric
stars. The stellar structure is obtained by solving Einstein's field
equation for a rotationally deformed 2D fluid
\cite{Weber,Glendenning2000}. The numerical method used here is based
on the KEH method \cite{Komatsu1989,Cook1992,Stergioulas1995}. The
equation of state used for computing the global neutron star
properties and its composition is a relativistic non-linear mean field
(RMF) model, whose parameters are adjusted to the properties of
nuclear matter at saturation density \cite{Glendenning1989}.

Once the structure of a rotating neutron star is computed, we solve
the equations that govern the thermal evolution. The latter equations
are re-derived for the metric of a 2D rotating neutron star.  We show
that the cooling of the object strongly depends on its frequency. At
higher frequencies, rotating neutron stars show a substantial
temperature difference between the equator and the pole. We also show
that the time scale for the thermalization of such stars increases
with its frequency.

This paper is organized as follows. In section \ref{Model} we discuss
the energy balance and transport equations for a rotating neutron
star. In section \ref{sec:surftemp} we show the results for the
surface temperature evolution of rotating neutron stars. The internal
temperature evolution is discussed in section \ref{sec:inttemp}.
Finally, in section \ref{sec:discussion} we present our conclusions.

\section{2D thermal evolution} \label{Model}

Given the complex nature of rotating neutron stars, where effects like
frame dragging (Lense-Thirring effect) and additional self-consistency
conditions (e.g.\ mass shedding, which sets an absolute limit on rapid
rotation) are in place, carrying out a 2D study of the thermal
evolution of such objects is a challenging task. First steps toward
this direction were made in
\cite{Schaab1998,Negreiros2011,Stejner2009,Pons2009, Cheng2004}.   We
note that the 2D cooling of magnetized neutron stars has been
  previously studied in \cite{Pons2009,Aguilera2008}, where the 2D
  effects of the magnetic field in the crust were investigated, and
  first steps towards a consistent magneto-thermal evolution
  simulation were taken.  The cooling simulations performed in
  \cite{Pons2009,Aguilera2008} are based on a standard spherically
  symmetric metric, since this study focused primarily on the impact
  of magnetic field effects on cooling.  Differently from
\cite{Pons2009}, we are presenting full 2D calculations of the thermal
evolution of rotating neutron stars, taking into account the 2D metric
of a rotating fluid distribution. The latter can be written as
\cite{Weber}
\begin{eqnarray}
ds^2 &=& - e^{2 \nu} dt^2 + e^{2 \phi} (d\varphi - N^\varphi dt)^2
\nonumber \\ &&+ e^{2 \omega} (dr^2 + r^2 d\theta^2) ,
\label{eq:3.1}
\end{eqnarray}
where $e^{2\phi} \equiv e^{2(\alpha + \beta)} r^2 \sin^2\theta$ and
$e^{2 \omega} \equiv e^{2(\alpha-\beta)}$.  The quantities $\nu$,
$\phi$ and $\omega$ denote metric functions, and $N^\varphi$ accounts
for the dragging of local inertial frames caused by the rotating
fluid. All these quantities are functions of $r$ and $\theta$, and are
implicitly dependent on $N^\varphi$. They need to be computed
self-consistently from Einstein's field equation, $G^{\bar\alpha
  \bar\beta} = 8 \pi T^{\bar\alpha \bar\beta}$, where $T^{\bar\alpha
  \bar\beta}$ denotes the fluid's energy momentum tensor.

The general relativistic equations of energy balance and transport are
derived from the condition of energy-momentum conservation, and can be
written as
\begin{eqnarray}
\partial_r \tilde H_{\bar r} + {1 \over r} \partial_\theta
\tilde H_{\bar\theta} = - r \, e^{\phi + 2\omega}
\left( {1 \over \Gamma} e^{2\nu} \epsilon + \Gamma C_V \partial_t 
\tilde T\right) \nonumber \\  - r \, \Gamma U e^{\nu+2\phi+\omega} 
\left( \partial_r \Omega + {1 \over r} \partial_\theta \Omega \right) ,
\label{eq:3.23} \\
\partial_r \tilde T = - {e^{\nu -\phi} \over{r \kappa}}  \tilde H_{\bar r}
- \Gamma^2 U e^{-\nu + \phi} \, \tilde T \partial_r \Omega \, ,
\label{eq:3.24} \\
{1 \over r} \partial_\theta \tilde T = - {e^{-\nu -\phi} \over{r \kappa}}
 \tilde H_{\bar \theta}
- \Gamma^2 U e^{-\nu + \phi} \, \tilde T {1 \over r} \partial_\theta \Omega ,
\label{eq:3.25} \\
\Gamma U \partial_t \tilde T = - {e^{-\omega -\phi} \over{r \kappa}}
 \tilde H_{\bar \varphi} \, ,
\label{eq:3.26}
\end{eqnarray} 
where $\tilde H_i \equiv r e^{2\nu+\phi+\omega} H_i / \Gamma$, with
$H_i$ being the i-th component of the heat flux; $\tilde T \equiv
e^\nu T / \Gamma$, with $T$ being the temperature; $\kappa$ is the
thermal conductivity; $C_V$ is the specific heat; $\epsilon$ is the
neutrino emissivity; and the Lorentz factor $\Gamma \equiv (1 -
U^2)^{-1/2}$, where $U$ is the the proper velocity with respect to a
zero angular momentum observer, given by $U = (\Omega -
N^\varphi)e^{\phi}$. In the case of rigid body rotation, which is
considered here, one has $\Omega=$const so that Eqs.\ (\ref{eq:3.23})
to (\ref{eq:3.26}) reduce to
\begin{eqnarray}
\partial_r \tilde H_{\bar r} + {1 \over r} \partial_\theta \tilde
H_{\bar\theta} = - r \, e^{\phi + 2(\alpha-\beta)} \left( {e^{2\nu}
  \over \Gamma} \epsilon + \Gamma C_V \partial_t \tilde T\right) ,
\label{eq:3.29} \\
\partial_r \tilde T = - {1 \over{r \kappa}} e^{-\nu -\phi} \tilde H_{\bar r},
\, \quad \quad \quad \quad \quad \quad \quad \quad
\label{eq:3.30} \\
{1 \over r} \partial_\theta \tilde T = - {1 \over{r \kappa}} e^{-\nu
  -\phi} \tilde H_{\bar \theta}. \, \quad \quad \quad \quad \quad
\quad \quad \quad
\label{eq:3.31} 
\end{eqnarray} 
The standard cooling equations of spherically symmetric, non-rotating
neutron stars are obtained from Eqs.\ (\ref{eq:3.29}) through
(\ref{eq:3.31}) for $\Omega= 0$ and $\partial_\theta \tilde T=0$
\cite{Weber}. This work aims at solving Eqs.\ (\ref{eq:3.29}) to
(\ref{eq:3.31}) for the temperature distribution $T(r,\theta;t)$ of
non-spherical, rotating neutron stars.  The boundary conditions are
obtained by defining $\tilde H_{\bar r}$ at $r=0$ and $R(\theta)$, and
the heat flux $\tilde H_{\bar \theta}$ at $\theta=0, \pi/2$ and at
$r=R(\theta)$, with $R(\theta)$ denoting the stellar radius. The
star's initial temperature, $T(r,\theta;t=0)$, is typically chosen as
$\tilde T \equiv 10^{11}$~K.  Equations (\ref{eq:3.24}) and
(\ref{eq:3.25}) can be solved for $\tilde H_{\bar r}$ and $\tilde
H_{\bar \theta}$, differentiated with respect to $r$ and $\theta$,
respectively, and then substituted into Eq.\ (\ref{eq:3.23}), which
leads to the following parabolic differential equation (for
$\triangledown\Omega =0$),
\begin{eqnarray}
  \partial_t \tilde T = - \, {e^{2 \nu \over{\Gamma^2}} } {\epsilon
    \over{ C_V}} + {1 \over{r^2 \sin\theta}} {e^{3\nu-\gamma-2\xi}
    \over \Gamma} {1 \over {C_V}}\times \quad \quad \quad \quad \quad
  \quad \nonumber & \\ \Bigl( \partial_r \left( r^2 \kappa \sin \theta
  \, e^\gamma \left( \partial_r \tilde T \right) \right) + {1
    \over{r^2}} \partial_\theta \left( r^2 \kappa \sin \theta \,
  e^\gamma \left( \partial_\theta \tilde T \right) \right) \Bigr) \, ,
  & \nonumber \\
\label{eq:3.49}
\end{eqnarray}
with the definitions $r \sin\theta e^{-\nu+\gamma} = e^\phi$ and
$e^{-\nu+\xi} = e^{\alpha-\beta}$. Equation (\ref{eq:3.49}) is solved
numerically in combination with a fully general relativistic rotation
code, which provides the necessary macroscopic input quantities.  The
microscopic input quantities (e.g.\ specific heat, thermal
conductivity, neutrino emissivity) are obtained from the chosen model
for equation of state.

\begin{figure}
 \centering
 \vspace{1.0cm}
 \includegraphics[width=8.cm]{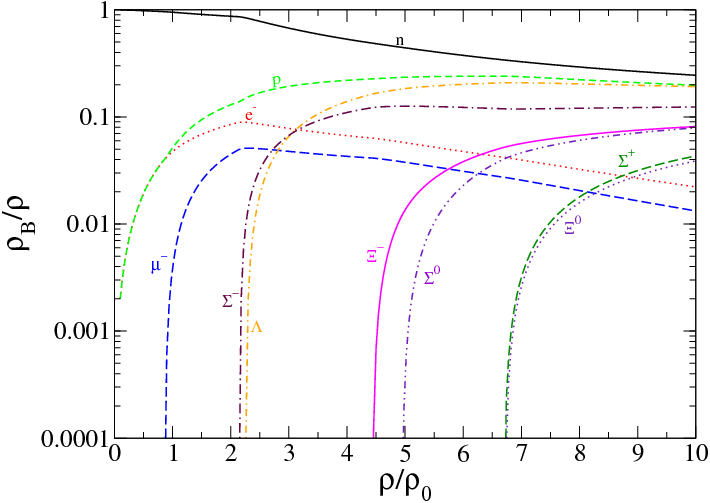}
 \caption{\label{part_pop}(Color online) Particle composition obtained
   from Eq.\ (\ref{lag}). The quantity $\rho$ denotes the baryon
   number density, $\rho_0$ is the number density at nuclear matter
   saturation density (0.16 fm$^{-3}$).}
 \end{figure}

We should devote some time to discuss the microscopic model for the
equation of state (EoS). In this work we used a non-linear
relativistic mean-field model (parameter set G300)
\cite{Glendenning1989}, which reproduces the properties of nuclear
matter at saturation density. The underlying nuclear lagrangian has
the form \cite{Weber,Glendenning2000,Glendenning1989}
\begin{equation}
  \begin{aligned}
    {\cal L} & = \sum_B \overline{\psi}_B [ \gamma_\mu ( i
      \partial^\mu -g_\omega \omega^\mu -g_\rho \vec{\rho}_\mu
      \overline{\psi} \gamma^\mu \vec{\tau} ) - (\nmass -g_\sigma
      \sigma) ]\psi_B \\ &+ \frac{1}{2}(\partial_\mu \sigma
    \partial^\mu \sigma - m_\sigma ^2 \sigma^2) - \frac{1}{3}b_\sigma
    \nmass (g_\sigma \sigma)^3 -\frac{1}{4}c_\sigma (g_\sigma
    \sigma)^4 \\ & -\frac{1}{4} \omega_{\mu \nu} \omega^{\mu \nu}
    +\frac{1}{2}m_\omega ^2 \omega_\mu \omega^\mu +\frac{1}{2}m_\rho
    ^2 \vec{\rho}_\mu \cdot \vec{\rho\,}^\mu -\frac{1}{4} \vec{\rho}_{\mu \nu}
    \cdot \vec{\rho\,}^{\mu \nu} \\ &+ \sum_{\lambda=e^-, \mu^-}
    \overline{\psi}_{\lambda} (i \gamma_\mu \partial ^\mu
    -m_{\lambda})\psi_{\lambda} \, ,
      \label{lag} 
      \end{aligned}
      \end{equation}
      where $B$ stands for protons ($p$), neutrons ($n$), and hyperons
      ($\Sigma, \Lambda, \Xi$).  The interactions among these
      particles are described via the exchange of $\sigma, \; \rho,\;
      \text{and} \; \omega$ mesons.  Their masses are
      $m_\sigma=550$~MeV, $m_\rho=769$~MeV, and $m_\omega=783$~MeV and
      their coupling constants are given by $g_\sigma = 9.1373$,
      $g_\rho = 8.3029$, and $g_\omega = 8.6324$. The coupling
      constants of the self-interaction term of the $\sigma$ meson are
      $b = 0.0033005$ and $c = 0.01529$ \cite{Glendenning2000}. The
      particle population of neutron star matter computed from Eq.\
      (\ref{lag}) is shown in Fig.\ \ref{part_pop}.

The G300 EoS leads to a relatively high proton fraction, triggering the
direct Urca process (DU hereafter) \cite{Lattimer1991} in neutron
stars with masses above $\sim 1.0$ M$_\odot$. As discussed in previous
studies
\cite{Schaab1996,Page2004,Page2006,Page2009,Blaschke2000,Grigorian2005,%
  Blaschke2006,Page2011,Yakovlev2011}, the presence of the DU process
enhances the cooling, leading to possible disagreement with observed
data \cite{Page2004,Page2009}. This issue can be resolved by assuming
that portions of the hadronic matter in the cores of neutron stars are
in a superfluid state \cite{Schaab1996,Levenfish1994,Page2011}.  The
presence of an enhanced cooling process allows us to track the heat
propagation more clearly inside the rotating neutron star. This is due
to the fact that, for rotating stars, the cold front originating from
the stellar core is more intense (due to stronger neutrino emission,
see \cite{Gnedin2001}), and thus easier to track. 

As for the neutrino emission processes taking place in the core, we
have considered the direct as well as the modified Urca processes
together with bremsstrahlung processes.  A detailed review of the
emissivities of such processes can be found in reference
\cite{Yakovlev2001a}. In addition to the core, we also consider the
standard processes that take place in the crust of a neutron star
\cite{Yakovlev2001a}. The specific heat of the hadrons is given by the
usual specific heat of fermions, as described in \cite{Page2004}.  For
the thermal conductivity we follow the calculations of
\cite{Flowers1981}.

  In the present study, we are not considering the possibility that
  neutrons and protons may be in a superfluid and superconducting
  state, respectively \cite{Weber,Page2006,Page2011,Yakovlev2011}. The
  reason being that
  one obtains somewhat sharper temperature gradients inside the
  neutron stars if the neutrino emission rates are not suppressed by
  superfluidity, which allows us to assess the effects of the 2D
  structure on the cooling of rotating neutron stars more clearly.
  The consequences of superfluidity/superconductivity, along with the
  Cooper-pair breaking/formation process, on the thermal evolution of
  rotating neutron stars will be presented in a follow-up study.

\section{Surface Temperature Evolution}\label{sec:surftemp}

We now present the results for the thermal evolution of rotating
neutron stars. We analyze the thermal evolution of three different
stellar configurations, each one with the same central density but
rotating at different selected frequencies,  ranging from around
  100~Hz to 800~Hz (see table \ref{table:stars}).

  The choice of this frequency range is motivated by theoretical
  calculations of the mapping between initial and final neutron star
  spins, likely spin-down mechanisms, and observational constraints
  from pulsars and supernova energetics. On this basis, it has been
  argued \cite{Ott2006} that the initial pre-collapse central iron
  core periods inside of massive stars, which give birth to neutron
  stars and pulsars, are on average greater than around 50
  seconds. The associated neutron stars would then be born with
  rotation periods greater than around 10 ms
  \cite{Ott2006,Kaspi2006}. Considering the many poorly understood
  features associated with the birth of rotating neutron stars, such
  as the estimated iron core spin rates, angular momentum profiles,
  progenitor masses, and general relativistic effects, the central
  iron core periods may be smaller by a factor of 5, however, in which
  case the birth periods of neutron stars increases to around 1000~Hz.
  Observationally, the most rapidly rotating neutron star, PSR
  J1748-2446ad, rotates at 716~Hz \cite{hessels06:a}, followed by the
  642~Hz pulsar B1937+21 \cite{backer82:a}. Very recently, it has been
  noted that the low mass X-ray binary SAX J1750.8-2900 may have a
  spin frequency of 601 Hz, and that this object may be the most
  luminous known neutron star \cite{Lowell2012}.

\begin{figure}
 \centering
 \vspace{1.0cm}
 \includegraphics[width=8.cm]{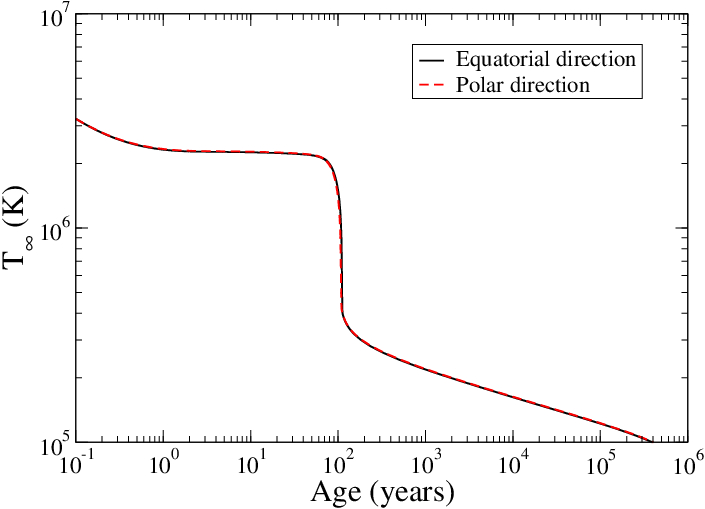}
 \caption{\label{Cool_1}(Color online) Redshifted temperature at
   star's pole and equator as a function of stellar age, for the 148
   Hz star of table \ref{table:stars}.}
 \end{figure}

\begin{figure}
 \centering
 \vspace{1.0cm}
 \includegraphics[width=8.cm]{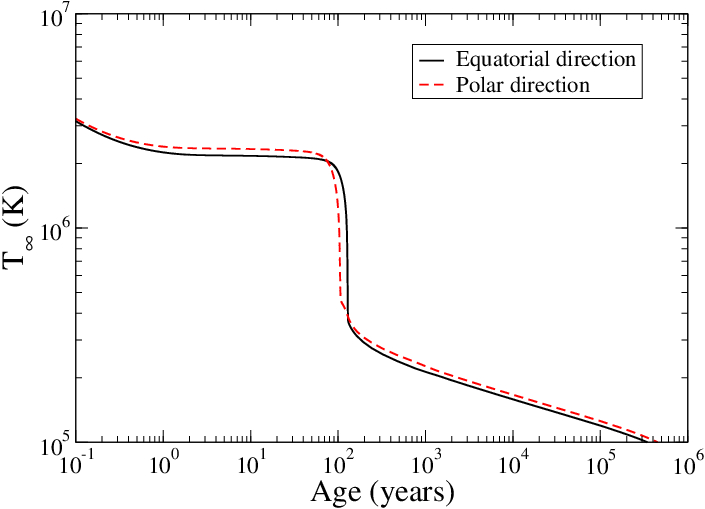}
 \caption{\label{Cool_2}(Color online) Same as Fig.~\ref{Cool_1}, but
   for the 488 Hz star of table \ref{table:stars}.}
 \end{figure}

\begin{figure}
 \centering
 \vspace{1.0cm}
 \includegraphics[width=8.cm]{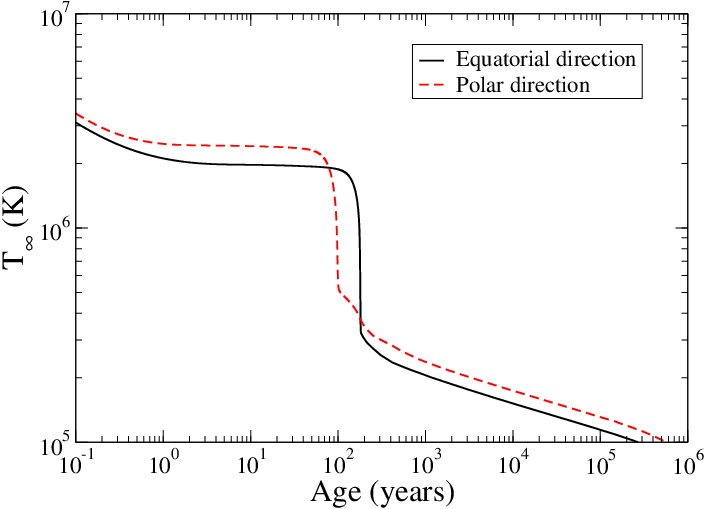}
 \caption{\label{Cool_3}(Color online) Same as Fig.~\ref{Cool_1}, but
   for the 755 Hz star of table \ref{table:stars}.}
 \end{figure}

Having said that, we now come back to the discussion of the properties
of the neutron stars listed in table \ref{table:stars}.
\begin{table}[h]
  \caption{\label{table:stars} Properties of neutron stars whose
    thermal evolution is studied in this paper. All stars have a
    central density of 350 MeV/fm$^3$. Their properties are computed
    for the relativistic mean-field EoS G300
    \cite{Glendenning1989}. $M$ denotes the gravitational mass, $R_e$
    the equatorial radius, $R_p$ the polar radius, $e =
    \sqrt{1-(R_p/R_e)^2}$ the eccentricity, $\Omega$ the star's
    rotational frequency, and $\Omega_K$ is the mass shedding frequency.}
\begin{ruledtabular}
\begin{tabular}{cccccc}
 $M/M_\odot$   & $R_e$ (km) & $R_p$ (km) & $e$ & $\Omega$ (Hz) & $\Omega_K$
(Hz) \\
\hline
       & &  & \\
  1.28 & 13.25  & 13.11 & 0.14 & 148 & 1330    \\ 
  1.34 & 13.85  & 12.49 & 0.43 & 488 & 1275    \\ 
  1.48 & 15.21  & 11.50 & 0.65 & 755 & 1169    \\ 
\end{tabular} 
\end{ruledtabular}
\end{table} 
The thermal cooling equation (\ref{eq:3.49}) was solved numerically
for each star in table \ref{table:stars}, and the results are shown in
Figs.~\ref{Cool_1}--\ref{Cool_3}, where the redshifted temperatures at
the poles and the equatorial belt of stars are plotted as a function
of time.
 
 In Figure~\ref{Cool_1} we show the cooling of a slowly rotating (148
 Hz) neutron star. At such low frequencies there is no significant
 difference between the polar and equatorial temperature evolution,
 indicating that such stars cool essentially like spherically
 symmetric stars (the eccentricity of this star is just 0.14, as shown
 in table \ref{table:stars}).  The situation is different for stars
 rotating at successively higher frequencies, as shown in
 Figs.~\ref{Cool_2} and \ref{Cool_3}. One sees that as the frequency
 increases, so does the difference between the polar and equatorial
 temperatures, with the pole being slightly warmer than the rest of
 the stellar surface due to the higher surface gravity there.

   It has been noted previously \cite{Aguilera2008,Pons2009} that
   anisotropic heat transport (caused by magnetic fields) also leads
   to the appearance of a hot spot in the magnetic pole. We note that
   in our case the hot spot is due to anisotropic heat transport due
   to rotation. Thus, if we were to include magnetic field effects, we
   ought to obtain an even more pronounced hot spot at the pole
   (assuming that the magnetic and rotation axis are either the same,
   or differ only by a small inclination angle). A study which
   explores the impact of both effects combined is currently under
   way.

 Another noticeable difference concerns to the sharp temperature drop
 at $\sim 100$ years, which happens simultaneously for the pole and
 equator of low frequency stars (i.e., spherical objects). This is not
 the case for stars rotating at higher frequencies, where a delay in
 the equatorial temperature drop is observed, as shown in
 Figs.~\ref{Cool_2} and \ref{Cool_3}.  One can understand this sudden
 drop in surface temperature as the moment in time where the "cold
 front", spreading from the core to the surface, has reached the
 stellar surface. (From this moment on the interior of the star is in
 thermal equilibrium.) If fast neutrino processes, like the direct
 Urca process, are active in the core, the cold front will be more
 distinctive, leading to a more pronounced drop in temperature. The
 time at which the cold front reaches the surface depends strongly on
 the properties of the crust \cite{Gnedin2001}. The cold front arrives
 first at the poles and then at the equator. This feature has its
 origin in the rotational deformation of the star: a flattening at the
 pole (leading to a thinner crust in this region) and an expansion at
 the equator (leading to a thicker crust). The thinner crust in the
 polar region, combined with the lower fluid velocity, leads to a more
 efficient heat propagation, allowing the cold front to reach the pole
 more quickly, as shown in Figs.\ \ref{Cool_2} and \ref{Cool_3}. This
 effect is less distinctive for stars with lower frequencies, since
 their shapes tend to be spherically symmetric and their crusts are
 more uniform.

\begin{figure}
 \centering
 \vspace{1.0cm}
 \includegraphics[width=8.cm]{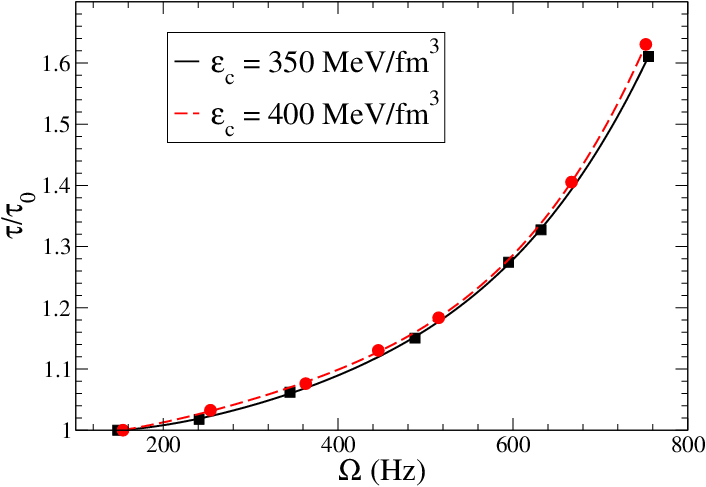}
 \caption{\label{tau_c_x_freq}(Color online) Dependence of core-crust
   thermal coupling time, $\tau$, on rotational frequency of a neutron
   star. The quantity $\epsilon_c$ denotes the star's central
   density. Solid dots show the outcome of our numerical calculations,
   the lines show a $4^{\rm th}$ order polynomial fit to the data.}
 \end{figure}

 Keeping in mind that the core-crust thermal coupling time ($\tau$)
 depends primarily on the crust properties, and that the macroscopic
 properties of the crust are directly connected to the stellar
 rotation rate, we now investigate the dependence of $\tau$ on the
 frequency of the star. In order to do that we calculate the cooling
 of two sets of stars, each set having central densities of 350 and
 400 MeV/fm$^3$, and covering the range of frequencies from 100 and
 800 Hz. For each star we calculate $\tau/\tau_0$, with $\tau_0$ being
 the core-crust coupling time in the spherically symmetric case (i.e.\
 $\Omega = 0$), and $\tau$ the corresponding value for rotating
 neutron stars (defined as the moment that the cold front reaches the
 equator).  The results are shown in Fig.~\ref{tau_c_x_freq}.  We find
 that $\tau/\tau_0$ can be fitted well by a $4^{\rm th}$ order polynomial,
 and that the result is independent of the microscopic properties of
 the core, as can be seen in Fig.~\ref{tau_c_x_freq}.  The indication
 is that this result should hold regardless of the EoS used for the
 core, as long as the crustal composition is the same.

 We note that in this study we have used the
Baym-Pethick-Sutherland (BPS) model for the crust. The neutrino
emission processes taking place in the crust are Bremsstrahlung,
$e^+e^-$ pair annihilation, and plasmon decay processes. A
comprehensive overview of the neutrino emission processes in NS can be
found in \cite{Weber,Yakovlev2001a}. We have also
assumed a non-magnetized neutron star atmosphere, composed of iron
and iron-like elements. In which case the relationship between the
mantle ($T_m$) and surface temperature ($T_s$) is given by the
traditional relationship \cite{Gudmundsson1983}
\begin{equation}
 T_{m8} = 1.288\times \left(\frac{T^4_{s6}}{g_{14}}\right)^{0.455},
\label{Ts}
\end{equation}
where $T_{m8}$ is in units of $10^8$ K, $T_{s6}$ in $10^6$ K, and
$g_{14}$
is the surface gravity in units of $10^{14}$ cm/s$^2$.

\section{Interior Temperature Evolution}\label{sec:inttemp}

In order to better understand the thermal evolution of a rotating
star, we now analyze the interior temperature evolution of such
objects. The star rotating at $\Omega = 755$ Hz will be used for this
analysis, since its strong deformation allows us to better evaluate
the effects of rotation on cooling. The results of this analysis can
be generalized to stars rotating at lower frequencies.
\begin{figure*}[h]
\begin{center}
\begin{tabular}{cc}
 \includegraphics[width=0.50\textwidth]{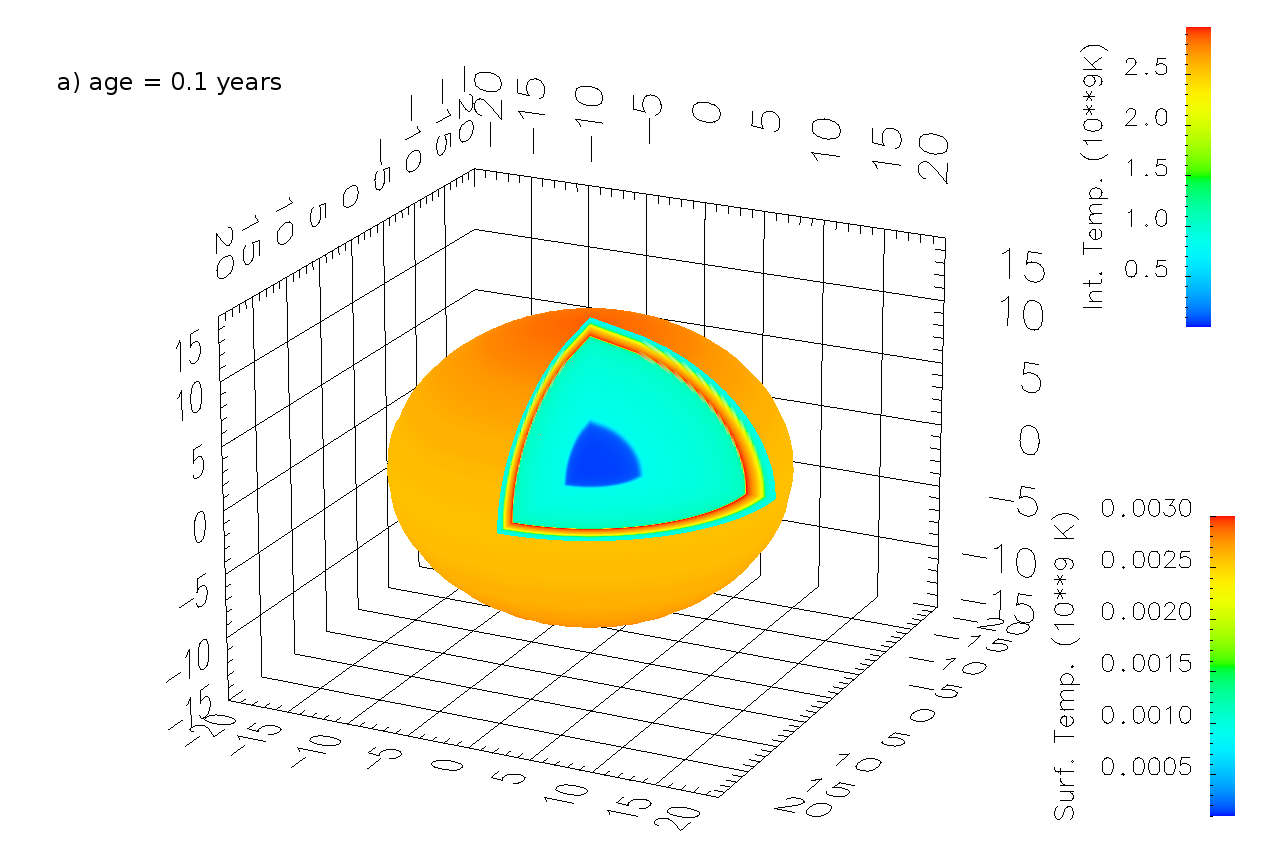} ~~&
 \includegraphics[width=0.50\textwidth]{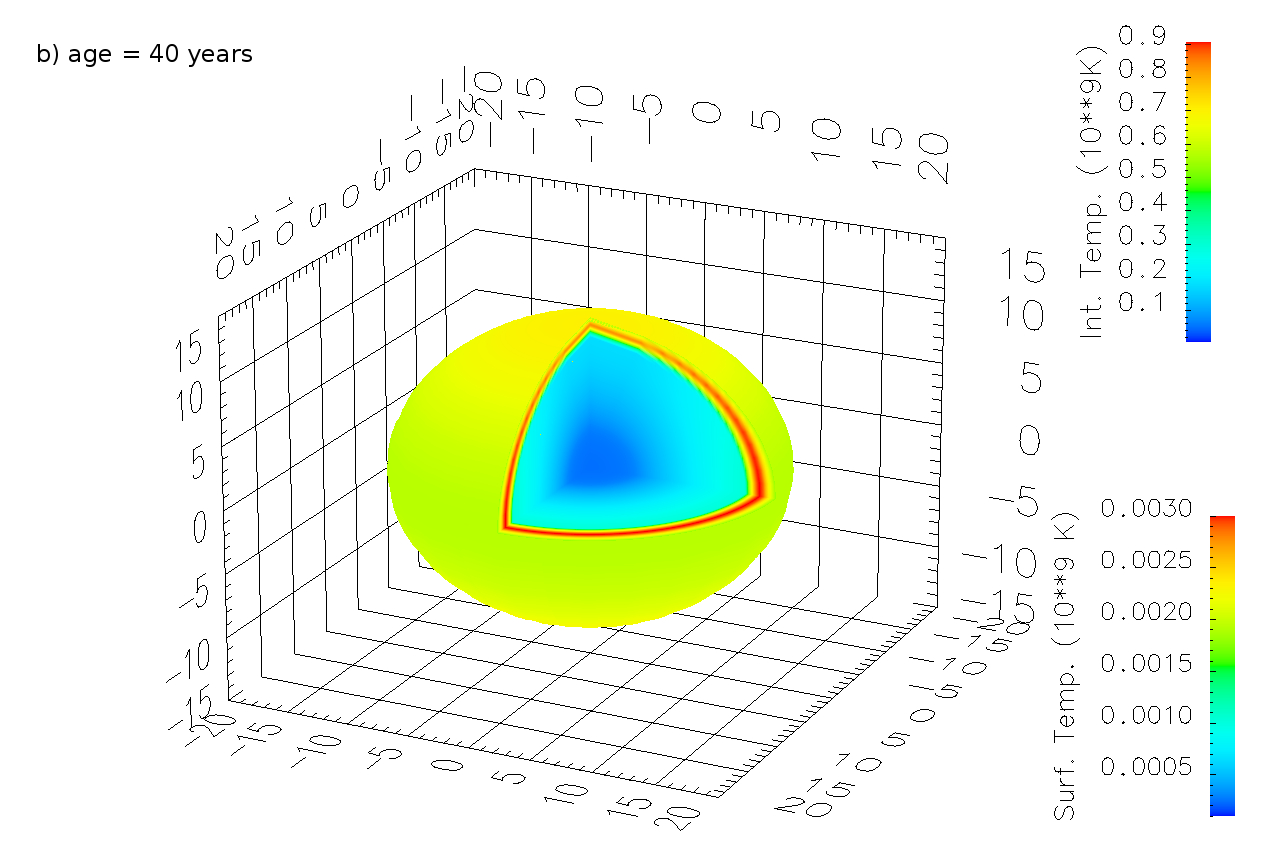} \\
& \\
 & \\
 & \\
\includegraphics[width=0.50\textwidth]{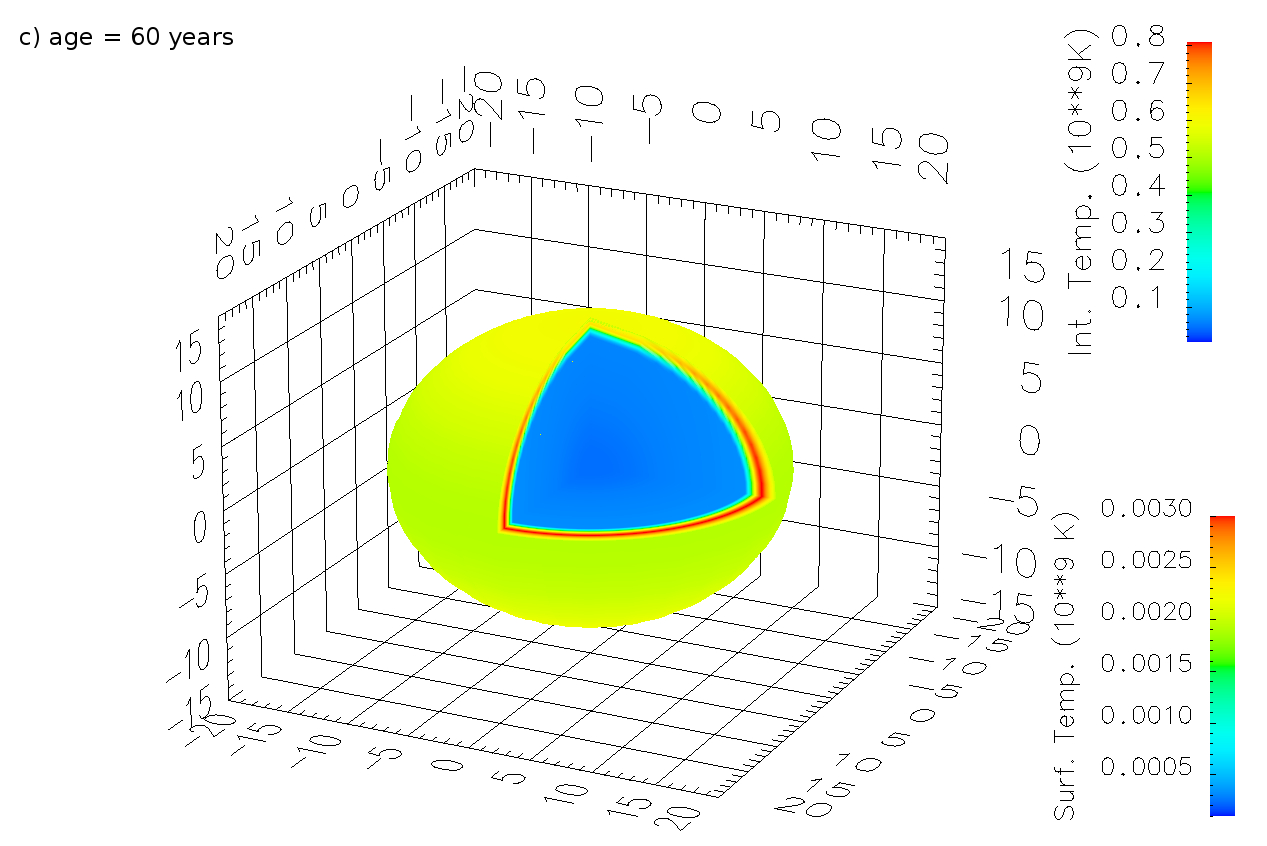} ~~&
\includegraphics[width=0.50\textwidth]{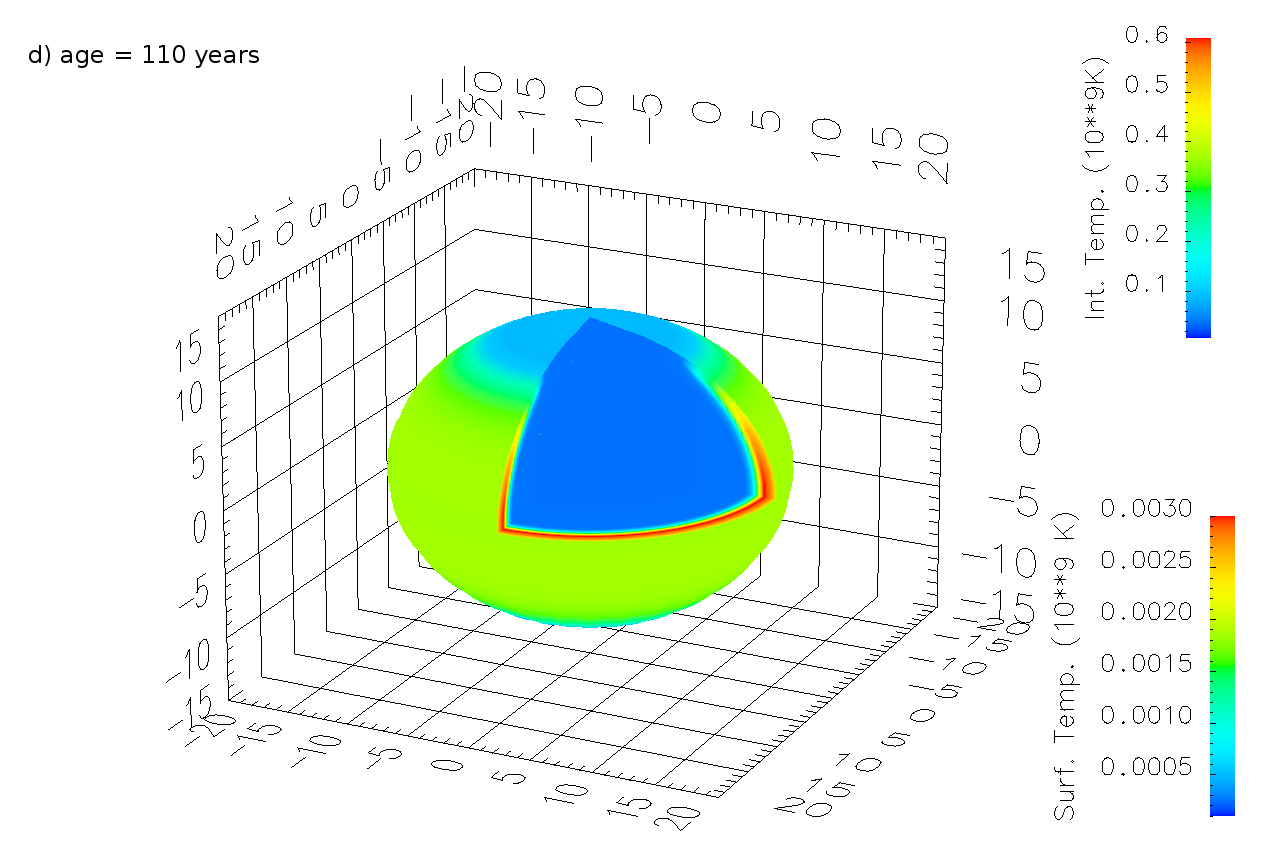}\\
& \\
 & \\
 & \\
\includegraphics[width=0.50\textwidth]{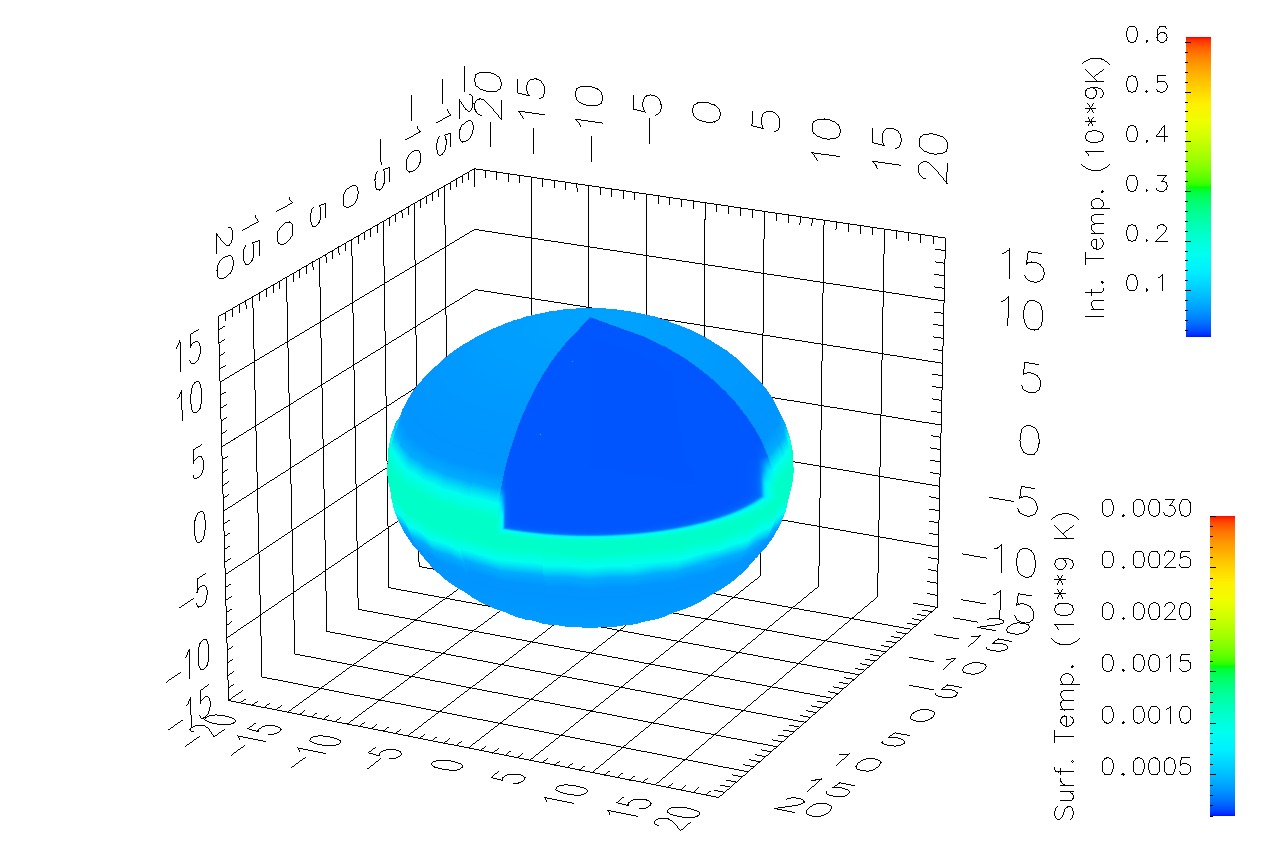} ~~&
\includegraphics[width=0.50\textwidth]{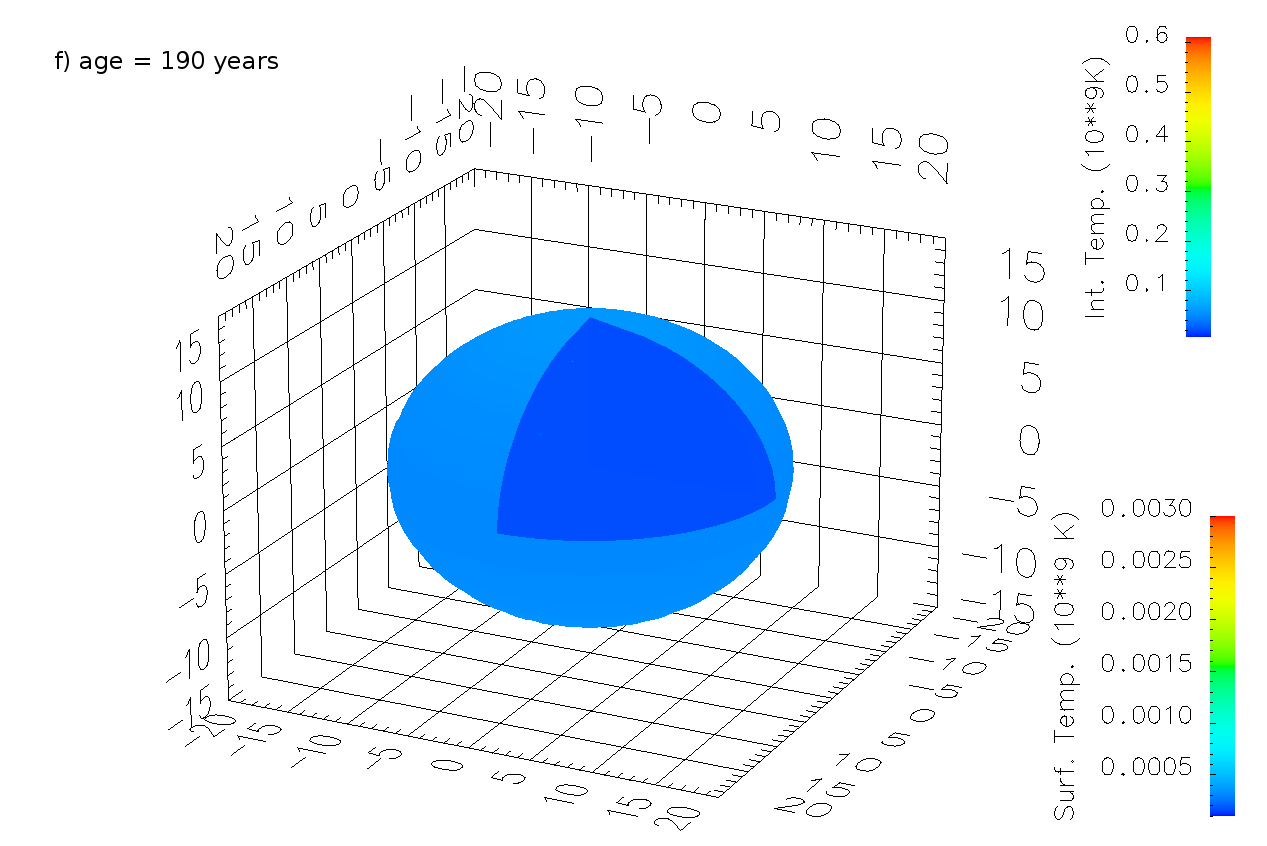}
\end{tabular}
\caption{(Color online) Temperature profiles of a neutron star
  rotating at $\Omega = 755$ Hz at different stages of stellar
  evolution, ranging from 0.1 (a), 40 (b), 60 (c), 110 (d), 170 (e),
  to 190 (f) years. The colors represent the redshifted
  temperature. The stars are cut open to help visualize the interior
  temperature evolutions. \label{T_int}}
\end{center}
\end{figure*}

Figure~\ref{T_int} shows a series of snapshots of the temperature
profiles of the $\Omega = 755$ Hz star. Displayed are different
cooling stages of this neutron star. For an early age of 0.1 year
(Fig.~\ref{T_int} (a)) the core and crust are thermally decoupled,
which can be clearly seen by the much higher temperature of the
crust. Also clearly noticeably is the higher temperature of the pole,
which is due to the greater surface gravity in this
region. Fig.~\ref{T_int} (a) also shows that the core is much colder
than the outer layers, which is a consequence of the fast core cooling
via the DU process. The cold interior region will propagate like a
``cold front'' until it finally reaches the surface, at which point
the star have become thermalized. The propagation of the cold front
can be observed in Fig.~\ref{T_int} (b), which shows the temperature
profile at a stellar age of 40 years. As already mentioned in section
\ref{sec:surftemp}, the cold front propagates more efficiently in the
polar direction, allowing this region to couple with the core more
quickly than the equatorial regime (see also Figs.~\ref{Cool_1} to
\ref{Cool_3}). Figure~\ref{T_int} (c), (d) and (e) illustrate stellar
epochs where the cold front is reaching the surface, first the polar
region (Fig.~\ref{T_int} (c)), and then propagating downward towards
the equatorial region (Fig.~\ref{T_int} (d) and (e)). After 190 years,
the stellar interior has reached thermal equilibrium, as shown in
Fig.~\ref{T_int} (f).  For ages greater than 190 years the pole is
warmer than the equator, as was the case for Figs.~\ref{T_int} (a) and
(b). The temperature difference is too small, however, to be
noticeable in Fig.~\ref{T_int} (f).  

The surface temperatures shown in Fig.~\ref{T_int} are the redshifted
surface temperatures. This temperature is a function of the mantle's
temperature as given by Eq.~(\ref{Ts})
\cite{Gudmundsson1983,Potekhin1997}.

\section{Discussion and Conclusions}\label{sec:discussion}

The objective of this work was is to investigate in detail the cooling
of rotationally deformed neutron stars in 2 dimensions (2D). Our
investigation is fully self-consistent. Einstein's field equations of
rotating compact objects have been solved in combination with a 2D
cooling code. Such studies are extremely important if one wants to
understand the thermal evolution of stellar systems where spherical
symmetry is broken. This might be the case for highly magnetized
neutron stars, neutron stars that are spinning down from high to low
rotation rates, as well as for accreting neutron stars which are being
spun up to high rotation rates.  This work represents the first step
towards a widely applicable, fully self-consistent treatment of the
cooling of neutron stars whose spherical symmetries have been
broken. As a first step in this direction, we considered here rigidly
rotating, rotationally deformed neutron stars. The metric of such
objects is significantly different from the metric of spherically
symmetric stars; we therefore re-derived the general relativistic
equations that govern the thermal evolution of such stars.

The microscopic model for the equation of state (EoS) which was used
in our study is that of purely hadronic matter, with the whole baryon
octet included. This model for the EoS allows for the direct Urca
process in stars with masses above 1.0~$\msun$, which leads to
enhanced stellar cooling. This choice for the equation of state was
intentional, since the direct Urca process yields sharper temperature
gradients inside the star, which allow us to track the energy
transport more clearly.  After obtaining a solid understanding of the
thermal evolution of rotating compact stars, our study will be applied
in a future study to compact objects with more complex core
compositions, such as hadronic superfluids, boson condensates, or
quark matter.

Our results indicate that rotation (and hence the 2D structure) plays
an important role for the thermal evolution of neutron stars. The most
important features can be summarized as follows:
\begin{itemize}
\item The redshifted temperature of the pole is slightly higher than
  the temperature at the equator, which is due the difference in
  surface gravity between these two regions.
 
\item Thermal energy is more efficiently transported along the polar
  direction, due to the lower fluid velocities and thinner crust
  there. This allows a ``cold front'', originating from the core, to
  arrive at the pole sooner than at the equator.

\end{itemize}
With respect to the second item above, we have found that the increase
in the core-crust coupling time is independent of all stellar
properties, except the frequency. This is consistent with the fact
that the crusts of the neutron star at different rotational stages are
very similar regardless of their core composition, and that the only
factor (in the present study) that alters the macroscopic properties
of the crust is rotation. This is indicated by the increase in the
core-crust coupling time as a function of frequency, which is very
similar for objects with different central densities (and thus
different particle compositions).

\bigskip

 {\bf Acknowledgement.}
F.W.\ is supported by the National Science Foundation (USA)
under Grant PHY-0854699. R.N.\ and S.S.\ acknowledge access to the
computer facilities of the CSC Frankfurt. R.N.\ acknowledges
financial support from the LOEWE program HIC for FAIR. 


\begin{thebibliography}{40}
\expandafter\ifx\csname natexlab\endcsname\relax\def\natexlab#1{#1}\fi
\expandafter\ifx\csname bibnamefont\endcsname\relax
  \def\bibnamefont#1{#1}\fi
\expandafter\ifx\csname bibfnamefont\endcsname\relax
  \def\bibfnamefont#1{#1}\fi
\expandafter\ifx\csname citenamefont\endcsname\relax
  \def\citenamefont#1{#1}\fi
\expandafter\ifx\csname url\endcsname\relax
  \def\url#1{\texttt{#1}}\fi
\expandafter\ifx\csname urlprefix\endcsname\relax\def\urlprefix{URL
}\fi
\providecommand{\bibinfo}[2]{#2}
\providecommand{\eprint}[2][]{\url{#2}}

\bibitem[{\citenamefont{Tsuruta and Cameron}(1965)}]{Tsuruta1965}
\bibinfo{author}{\bibfnamefont{S.}~\bibnamefont{Tsuruta}},
  \bibinfo{journal}{Nature} \textbf{\bibinfo{volume}{207}},
  \bibinfo{pages}{364} (\bibinfo{year}{1965}).

\bibitem[{\citenamefont{Maxwell}(1979)}]{Maxwell1979}
\bibinfo{author}{\bibfnamefont{O.~V.} \bibnamefont{Maxwell}},
  \bibinfo{journal}{The Astrophysical Journal}
\textbf{\bibinfo{volume}{231}},
  \bibinfo{pages}{201} (\bibinfo{year}{1979}).

\bibitem[{\citenamefont{Schaab et~al.}(1996)\citenamefont{Schaab,
Weber,
  Weigel, and Glendenning}}]{Schaab1996}
\bibinfo{author}{\bibfnamefont{C.}~\bibnamefont{Schaab}},
  \bibinfo{author}{\bibfnamefont{F.}~\bibnamefont{Weber}},
  \bibinfo{author}{\bibfnamefont{M.~K.}~\bibnamefont{Weigel}},
\bibnamefont{and}
  \bibinfo{author}{\bibfnamefont{N.~K.} \bibnamefont{Glendenning}},
  \bibinfo{journal}{Nuclear Phys A} \textbf{\bibinfo{volume}{605}},
  \bibinfo{pages}{531} (\bibinfo{year}{1996}).

\bibitem[{\citenamefont{Page et~al.}(2004)\citenamefont{Page,
Lattimer,
  Prakash, and Steiner}}]{Page2004}
\bibinfo{author}{\bibfnamefont{D.}~\bibnamefont{Page}},
  \bibinfo{author}{\bibfnamefont{J.}~\bibnamefont{Lattimer}},
  \bibinfo{author}{\bibfnamefont{M.}~\bibnamefont{Prakash}},
\bibnamefont{and}
  \bibinfo{author}{\bibfnamefont{A.~W.} \bibnamefont{Steiner}},
  \bibinfo{journal}{The Astrophysical Journal Supplement Series}
  \textbf{\bibinfo{volume}{155}}, \bibinfo{pages}{623}
(\bibinfo{year}{2004}).

\bibitem[{\citenamefont{Page et~al.}(2006)\citenamefont{Page, Geppert,
and
  Weber}}]{Page2006}
\bibinfo{author}{\bibfnamefont{D.}~\bibnamefont{Page}},
  \bibinfo{author}{\bibfnamefont{U.}~\bibnamefont{Geppert}},
\bibnamefont{and}
  \bibinfo{author}{\bibfnamefont{F.}~\bibnamefont{Weber}},
  \bibinfo{journal}{Nuclear Physics A} \textbf{\bibinfo{volume}{777}},
  \bibinfo{pages}{497} (\bibinfo{year}{2006}).

\bibitem[{\citenamefont{Page et~al.}(2009)\citenamefont{Page,
Lattimer,
  Prakash, and Steiner}}]{Page2009}
\bibinfo{author}{\bibfnamefont{D.}~\bibnamefont{Page}},
  \bibinfo{author}{\bibfnamefont{J.}~\bibnamefont{Lattimer}},
  \bibinfo{author}{\bibfnamefont{M.}~\bibnamefont{Prakash}},
\bibnamefont{and}
  \bibinfo{author}{\bibfnamefont{A.~W.} \bibnamefont{Steiner}},
  \bibinfo{journal}{The Astrophysical Journal}
\textbf{\bibinfo{volume}{707}},
  \bibinfo{pages}{1131} (\bibinfo{year}{2009}).

\bibitem[{\citenamefont{Blaschke et~al.}(2000)\citenamefont{Blaschke,
Klahn,
  and Voskresensky}}]{Blaschke2000}
\bibinfo{author}{\bibfnamefont{D.}~\bibnamefont{Blaschke}},
  \bibinfo{author}{\bibfnamefont{T.}~\bibnamefont{Klahn}},
\bibnamefont{and}
  \bibinfo{author}{\bibfnamefont{D.~N.}~\bibnamefont{Voskresensky}},
  \bibinfo{journal}{The Astrophysical Journal}
\textbf{\bibinfo{volume}{533}},
  \bibinfo{pages}{406} (\bibinfo{year}{2000}).

\bibitem[{\citenamefont{Grigorian
et~al.}(2005)\citenamefont{Grigorian,
  Blaschke, and Voskresensky}}]{Grigorian2005}
\bibinfo{author}{\bibfnamefont{H.}~\bibnamefont{Grigorian}},
  \bibinfo{author}{\bibfnamefont{D.}~\bibnamefont{Blaschke}},
\bibnamefont{and}
  \bibinfo{author}{\bibfnamefont{D.}~\bibnamefont{Voskresensky}},
  \bibinfo{journal}{Physical Review C} \textbf{\bibinfo{volume}{71}},
  \bibinfo{pages}{045801} (\bibinfo{year}{2005}).

\bibitem[{\citenamefont{Blaschke et~al.}(2006)\citenamefont{Blaschke,
  Voskresensky, and Grigorian}}]{Blaschke2006}
\bibinfo{author}{\bibfnamefont{D.}~\bibnamefont{Blaschke}},
  \bibinfo{author}{\bibfnamefont{D.}~\bibnamefont{Voskresensky}},
  \bibnamefont{and}
  \bibinfo{author}{\bibfnamefont{H.}~\bibnamefont{Grigorian}},
  \bibinfo{journal}{Nuclear Physics A} \textbf{\bibinfo{volume}{774}},
  \bibinfo{pages}{815} (\bibinfo{year}{2006}).

\bibitem[{\citenamefont{Page
et~al.}(2011{\natexlab{a}})\citenamefont{Page,
  Prakash, Lattimer, and Steiner}}]{Page2011}
\bibinfo{author}{\bibfnamefont{D.}~\bibnamefont{Page}},
  \bibinfo{author}{\bibfnamefont{M.}~\bibnamefont{Prakash}},
  \bibinfo{author}{\bibfnamefont{J.}~\bibnamefont{Lattimer}},
\bibnamefont{and}
  \bibinfo{author}{\bibfnamefont{A.~W.}~\bibnamefont{Steiner}},
  \bibinfo{journal}{Physical Review Letters}
\textbf{\bibinfo{volume}{106}},
  \bibinfo{pages}{081101} (\bibinfo{year}{2011}{\natexlab{a}}).

\bibitem[{\citenamefont{Yakovlev et~al.}(2011)\citenamefont{Yakovlev,
Ho,
  Shternin, Heinke, and Potekhin}}]{Yakovlev2011}
\bibinfo{author}{\bibfnamefont{D.~G.} \bibnamefont{Yakovlev}},
  \bibinfo{author}{\bibfnamefont{W.~C.~G.} \bibnamefont{Ho}},
  \bibinfo{author}{\bibfnamefont{P.~S.} \bibnamefont{Shternin}},
  \bibinfo{author}{\bibfnamefont{C.~O.} \bibnamefont{Heinke}},
  \bibnamefont{and} \bibinfo{author}{\bibfnamefont{A.~Y.}
  \bibnamefont{Potekhin}}, \bibinfo{journal}{Monthly Notices of the
Royal
  Astronomical Society} \textbf{\bibinfo{volume}{411}},
\bibinfo{pages}{1977}
  (\bibinfo{year}{2011}).

\bibitem[{\citenamefont{Schaab et~al.}(1997)\citenamefont{Schaab,
Voskresensky,
  Sedrakian, and F}}]{Voskresensky1997}
\bibinfo{author}{\bibfnamefont{C.}~\bibnamefont{Schaab}},
  \bibinfo{author}{\bibfnamefont{D.}~\bibnamefont{Voskresensky}},
  \bibinfo{author}{\bibfnamefont{A.}~\bibnamefont{Sedrakian}},
  \bibnamefont{and} \bibinfo{author}{\bibnamefont{F}},
  \bibinfo{journal}{Astronomy and Astrophysics}
\textbf{\bibinfo{volume}{604}},
  \bibinfo{pages}{591} (\bibinfo{year}{1997}).

\bibitem[{\citenamefont{Shternin et~al.}(2011)\citenamefont{Shternin,
Yakovlev,
  Heinke, Ho, and Patnaude}}]{Shternin2011}
\bibinfo{author}{\bibfnamefont{P.~S.} \bibnamefont{Shternin}},
  \bibinfo{author}{\bibfnamefont{D.~G.} \bibnamefont{Yakovlev}},
  \bibinfo{author}{\bibfnamefont{C.~O.} \bibnamefont{Heinke}},
  \bibinfo{author}{\bibfnamefont{W.~C.~G.} \bibnamefont{Ho}},
\bibnamefont{and}
  \bibinfo{author}{\bibfnamefont{D.~J.} \bibnamefont{Patnaude}},
  \bibinfo{journal}{Monthly Notices of the Royal Astronomical Society:
Letters}
  \textbf{\bibinfo{volume}{412}}, \bibinfo{pages}{L108}
(\bibinfo{year}{2011}).

\bibitem[{\citenamefont{Shovkovy and Ellis}(2002)}]{Shovkovy2002}
\bibinfo{author}{\bibfnamefont{I.~A.}~\bibnamefont{Shovkovy}}
\bibnamefont{and}
  \bibinfo{author}{\bibfnamefont{P.~J.}~\bibnamefont{Ellis}},
  \bibinfo{journal}{Physical Review C} \textbf{\bibinfo{volume}{66}},
  \bibinfo{pages}{015802} (\bibinfo{year}{2002}).

\bibitem[{\citenamefont{Prakash et~al.}(1992)\citenamefont{Prakash,
Lattimer,
  and Pethick}}]{Prakash1992}
\bibinfo{author}{\bibfnamefont{M.}~\bibnamefont{Prakash}},
  \bibinfo{author}{\bibfnamefont{J.}~\bibnamefont{Lattimer}},
\bibnamefont{and}
  \bibinfo{author}{\bibfnamefont{C.}~\bibnamefont{Pethick}},
  \bibinfo{journal}{Astrophysical Journal}
\textbf{\bibinfo{volume}{390}},
  \bibinfo{pages}{L77} (\bibinfo{year}{1992}).

\bibitem[{\citenamefont{Lattimer et~al.}(1994)\citenamefont{Lattimer,
van
  Riper, Prakash, and Prakash}}]{Lattimer1994}
\bibinfo{author}{\bibfnamefont{J.~M.} \bibnamefont{Lattimer}},
  \bibinfo{author}{\bibfnamefont{K.~A.} \bibnamefont{van Riper}},
  \bibinfo{author}{\bibfnamefont{M.}~\bibnamefont{Prakash}},
\bibnamefont{and}
  \bibinfo{author}{\bibfnamefont{M.}~\bibnamefont{Prakash}},
  \bibinfo{journal}{The Astrophysical Journal}
\textbf{\bibinfo{volume}{425}},
  \bibinfo{pages}{802} (\bibinfo{year}{1994}).

\bibitem[{\citenamefont{Gusakov et~al.}(2005)\citenamefont{Gusakov,
Kaminker,
  Yakovlev, and Gnedin}}]{Gusakov2005}
\bibinfo{author}{\bibfnamefont{M.~E.} \bibnamefont{Gusakov}},
  \bibinfo{author}{\bibfnamefont{A.~D.} \bibnamefont{Kaminker}},
  \bibinfo{author}{\bibfnamefont{D.~G.} \bibnamefont{Yakovlev}},
  \bibnamefont{and} \bibinfo{author}{\bibfnamefont{O.~Y.}
  \bibnamefont{Gnedin}}, \bibinfo{journal}{Monthly Notices of the
Royal
  Astronomical Society} \textbf{\bibinfo{volume}{363}},
\bibinfo{pages}{555}
  (\bibinfo{year}{2005}).

\bibitem[{\citenamefont{Horvath et~al.}(1991)\citenamefont{Horvath,
Benvenuto,
  and Vucetich}}]{Horvath1991}
\bibinfo{author}{\bibfnamefont{J.}~\bibnamefont{Horvath}},
  \bibinfo{author}{\bibfnamefont{O.}~\bibnamefont{Benvenuto}},
  \bibnamefont{and}
\bibinfo{author}{\bibfnamefont{H.}~\bibnamefont{Vucetich}},
  \bibinfo{journal}{Physical Review D} \textbf{\bibinfo{volume}{44}},
  \bibinfo{pages}{3797} (\bibinfo{year}{1991}).

\bibitem[{\citenamefont{Weber et~al.}(2009)\citenamefont{Weber,
Negreiros, and
  Rosenfield}}]{Weber2007b}
\bibinfo{author}{\bibfnamefont{F.}~\bibnamefont{Weber}},
  \bibinfo{author}{\bibfnamefont{R.}~\bibnamefont{Negreiros}},
  \bibnamefont{and}
  \bibinfo{author}{\bibfnamefont{P.}~\bibnamefont{Rosenfield}},
  \bibinfo{journal}{Astrophysics and Space Science Library}
  \textbf{\bibinfo{volume}{357}}, \bibinfo{pages}{213}
(\bibinfo{year}{2009}).

\bibitem[{\citenamefont{Negreiros
et~al.}(2010)\citenamefont{Negreiros,
  Dexheimer, and Schramm}}]{Negreiros2010}
\bibinfo{author}{\bibfnamefont{R.}~\bibnamefont{Negreiros}},
  \bibinfo{author}{\bibfnamefont{V.~A.} \bibnamefont{Dexheimer}},
  \bibnamefont{and}
\bibinfo{author}{\bibfnamefont{S.}~\bibnamefont{Schramm}},
  \bibinfo{journal}{Physical Review C} \textbf{\bibinfo{volume}{82}},
  \bibinfo{pages}{035803} (\bibinfo{year}{2010}).

\bibitem[{\citenamefont{Niebergal
et~al.}(2010)\citenamefont{Niebergal, Ouyed,
  Negreiros, and Weber}}]{Niebergal}
\bibinfo{author}{\bibfnamefont{B.}~\bibnamefont{Niebergal}},
  \bibinfo{author}{\bibfnamefont{R.}~\bibnamefont{Ouyed}},
  \bibinfo{author}{\bibfnamefont{R.}~\bibnamefont{Negreiros}},
  \bibnamefont{and}
\bibinfo{author}{\bibfnamefont{F.}~\bibnamefont{Weber}},
  \bibinfo{journal}{Physical Review D} \textbf{\bibinfo{volume}{81}},
  \bibinfo{pages}{043005} (\bibinfo{year}{2010}).

\bibitem[{\citenamefont{Alford et~al.}(2005)\citenamefont{Alford,
Jotwani,
  Kouvaris, Kundu, and Rajagopal}}]{Alford2005a}
\bibinfo{author}{\bibfnamefont{M.}~\bibnamefont{Alford}},
  \bibinfo{author}{\bibfnamefont{P.}~\bibnamefont{Jotwani}},
  \bibinfo{author}{\bibfnamefont{C.}~\bibnamefont{Kouvaris}},
  \bibinfo{author}{\bibfnamefont{J.}~\bibnamefont{Kundu}},
\bibnamefont{and}
  \bibinfo{author}{\bibfnamefont{K.}~\bibnamefont{Rajagopal}},
  \bibinfo{journal}{Physical Review D} \textbf{\bibinfo{volume}{71}},
  \bibinfo{pages}{114011} (\bibinfo{year}{2005}).

\bibitem[{\citenamefont{Heinke and Ho}(2010)}]{Heinke2010}
\bibinfo{author}{\bibfnamefont{C.~O.} \bibnamefont{Heinke}}
\bibnamefont{and}
  \bibinfo{author}{\bibfnamefont{W.~C.~G.} \bibnamefont{Ho}},
  \bibinfo{journal}{The Astrophysical Journal}
\textbf{\bibinfo{volume}{719}},
  \bibinfo{pages}{L167} (\bibinfo{year}{2010}).

\bibitem{Blaschke2011} D. Blaschke, H. Grigorian, D. N. Voskresensky,
  and F. Weber, {\tt arXiv:1108.4125v1 [nucl-th]}.

\bibitem{Negreiros2011} R. Negreiros, S. Schramm, and F. Weber,
{\tt arXiv:1103.3870v4 [astro-ph.HE]}.


\bibitem[{\citenamefont{Weber}(1999)}]{Weber}
\bibinfo{author}{\bibfnamefont{F.}~\bibnamefont{Weber}},
  {\it {Pulsars as astrophysical laboratories for nuclear and
  particle physics}} (\bibinfo{publisher}{Institute of Physics},
  \bibinfo{address}{Bristol}, \bibinfo{year}{1999}),
\bibinfo{edition}{1st} ed.

\bibitem[{\citenamefont{Glendenning}(2000)}]{Glendenning2000}
\bibinfo{author}{\bibfnamefont{N.~K.} \bibnamefont{Glendenning}},
  {\bibinfo{title}{ \it {Compact stars: nuclear physics, particle
physics, and
  general relativity}}} (\bibinfo{publisher}{Springer},
\bibinfo{year}{2000}),
  \bibinfo{edition}{1st} ed.

\bibitem[{\citenamefont{Komatsu et~al.}(1989)\citenamefont{Komatsu,
Eriguchi,
  and Hachisu}}]{Komatsu1989}
\bibinfo{author}{\bibfnamefont{H.}~\bibnamefont{Komatsu}},
  \bibinfo{author}{\bibfnamefont{Y.}~\bibnamefont{Eriguchi}},
\bibnamefont{and}
  \bibinfo{author}{\bibfnamefont{I.}~\bibnamefont{Hachisu}},
  \bibinfo{journal}{Royal Astronomical Society, Monthly Notices }
\textbf{\bibinfo{volume}{237}},
\bibinfo{pages}{355} (\bibinfo{year}{1989}).

\bibitem[{\citenamefont{Cook et~al.}(1992)\citenamefont{Cook, Shapiro,
and
  Teukolsky}}]{Cook1992}
\bibinfo{author}{\bibfnamefont{G.~B.} \bibnamefont{Cook}},
  \bibinfo{author}{\bibfnamefont{S.~L.} \bibnamefont{Shapiro}},
  \bibnamefont{and} \bibinfo{author}{\bibfnamefont{S.~A.}
  \bibnamefont{Teukolsky}}, \bibinfo{journal}{The Astrophysical
Journal}
  \textbf{\bibinfo{volume}{398}}, \bibinfo{pages}{203}
(\bibinfo{year}{1992}).

\bibitem[{\citenamefont{Stergioulas and
Friedman}(1995)}]{Stergioulas1995}
\bibinfo{author}{\bibfnamefont{N.}~\bibnamefont{Stergioulas}}
\bibnamefont{and}
  \bibinfo{author}{\bibfnamefont{J.~L.} \bibnamefont{Friedman}},
  \bibinfo{journal}{The Astrophysical Journal}
\textbf{\bibinfo{volume}{444}},
  \bibinfo{pages}{306} (\bibinfo{year}{1995}).

\bibitem[{\citenamefont{Glendenning}(1989)}]{Glendenning1989}
\bibinfo{author}{\bibfnamefont{N.~K.} \bibnamefont{Glendenning}},
  \bibinfo{journal}{Nuclear Physics A} \textbf{\bibinfo{volume}{493}},
  \bibinfo{pages}{521} (\bibinfo{year}{1989}).

\bibitem{Schaab1998} C. Schaab and M. K. Weigel, Astronomy and
  Astrophysics {\bf 336}, L13 (1998).

\bibitem[{\citenamefont{Stejner et~al.}(2009)\citenamefont{Stejner,
Weber, and
  Madsen}}]{Stejner2009}
\bibinfo{author}{\bibfnamefont{M.}~\bibnamefont{Stejner}},
  \bibinfo{author}{\bibfnamefont{F.}~\bibnamefont{Weber}},
\bibnamefont{and}
  \bibinfo{author}{\bibfnamefont{J.}~\bibnamefont{Madsen}},
  \bibinfo{journal}{The Astrophysical Journal}
\textbf{\bibinfo{volume}{694}},
  \bibinfo{pages}{1019} (\bibinfo{year}{2009}).


\bibitem{Cheng2004}
C. Y. Hui, and K. S. Cheng, The  Astrophysical Journal {\bf 608},
935 (2004).

\bibitem[{\citenamefont{Pons et~al.}(2009)\citenamefont{Pons,
Miralles, and
  Geppert}}]{Pons2009}
\bibinfo{author}{\bibfnamefont{J.~A.} \bibnamefont{Pons}},
  \bibinfo{author}{\bibfnamefont{J.~A.} \bibnamefont{Miralles}},
  \bibnamefont{and}
\bibinfo{author}{\bibfnamefont{U.}~\bibnamefont{Geppert}},
  \bibinfo{journal}{Astronomy and Astrophysics}
\textbf{\bibinfo{volume}{496}},
  \bibinfo{pages}{207} (\bibinfo{year}{2009}).


\bibitem{Aguilera2008}
D.~N. Aguilera, J.~A. Pons, and J.~A. Miralles, Astronomy and
Astrophysics {\bf 486}, 255 (2008).

\bibitem[{\citenamefont{Lattimer et~al.}(1991)\citenamefont{Lattimer,
Pethick,
  Prakash, and Haensel}}]{Lattimer1991}
\bibinfo{author}{\bibfnamefont{J.~M}~\bibnamefont{Lattimer}},
  \bibinfo{author}{\bibfnamefont{C.}~\bibnamefont{Pethick}},
  \bibinfo{author}{\bibfnamefont{M.}~\bibnamefont{Prakash}},
\bibnamefont{and}
  \bibinfo{author}{\bibfnamefont{P.}~\bibnamefont{Haensel}},
  \bibinfo{journal}{Physical Review Letters}
\textbf{\bibinfo{volume}{66}},
  \bibinfo{pages}{2701} (\bibinfo{year}{1991}).

\bibitem[{\citenamefont{Levenfish and Yakovlev}(1994)}]{Levenfish1994}
\bibinfo{author}{\bibfnamefont{K.~P.} \bibnamefont{Levenfish}}
  \bibnamefont{and} \bibinfo{author}{\bibfnamefont{D.~G.}
  \bibnamefont{Yakovlev}}, \bibinfo{journal}{Astronomy Letters}
  \textbf{\bibinfo{volume}{20}}, \bibinfo{pages}{43}
(\bibinfo{year}{1994}).


\bibitem[{\citenamefont{Gnedin et~al.}(2001)\citenamefont{Gnedin,
Yakovlev, and
  Potekhin}}]{Gnedin2001}
\bibinfo{author}{\bibfnamefont{O.~Y.} \bibnamefont{Gnedin}},
  \bibinfo{author}{\bibfnamefont{D.~G.} \bibnamefont{Yakovlev}},
  \bibnamefont{and} \bibinfo{author}{\bibfnamefont{A.~Y.}
  \bibnamefont{Potekhin}}, \bibinfo{journal}{Monthly Notices of the
Royal
  Astronomical Society} \textbf{\bibinfo{volume}{324}},
\bibinfo{pages}{725}
  (\bibinfo{year}{2001}).

\bibitem[{\citenamefont{Yakovlev et~al.}(2001)\citenamefont{Yakovlev,
Kaminker,
  Gnedin, and Haensel}}]{Yakovlev2001a}
\bibinfo{author}{\bibfnamefont{D.~G.} \bibnamefont{Yakovlev}},
  \bibinfo{author}{\bibfnamefont{A.~D.} \bibnamefont{Kaminker}},
  \bibinfo{author}{\bibfnamefont{O.~Y.} \bibnamefont{Gnedin}},
  \bibnamefont{and}
\bibinfo{author}{\bibfnamefont{P.}~\bibnamefont{Haensel}},
  \bibinfo{journal}{Physics Reports} \textbf{\bibinfo{volume}{354}},
  \bibinfo{pages}{1} (\bibinfo{year}{2001}).

\bibitem[{\citenamefont{Flowers and Itoh}(1981)}]{Flowers1981}
\bibinfo{author}{\bibfnamefont{E.}~\bibnamefont{Flowers}}
\bibnamefont{and}
  \bibinfo{author}{\bibfnamefont{N.}~\bibnamefont{Itoh}},
\bibinfo{journal}{The
  Astrophysical Journal} \textbf{\bibinfo{volume}{250}},
\bibinfo{pages}{750}
  (\bibinfo{year}{1981}).

\bibitem{Ott2006}
C. Ott, A. Burrows, T.A. Thompson, E. Livne, and R.
Walder, The Astrophysical Journal Supplement Series {\bf 164}, 130
(2006).

\bibitem{Kaspi2006}
C. A. Faucher-Giguere and V.M. Kaspi, The Astrophysical Journal {\bf
643}, 332 (2006).

\bibitem{hessels06:a} J. W. Hessels, S. M. Ransom, I. H. Stairs,
  P. C. Freire, V. M.  Kaspi, and F. Camilo, Science {\bf 311} (No.\ 5769)
  (2006) 1901.
 
\bibitem{backer82:a} D. C. Backer, S. R. Kulkarni, C. Heiles,
  M. M. Davis, and W. M. Goss, Nature {\bf 300} (1982) 615.

\bibitem{Lowell2012}
A.W. Lowell, J.A. Tomsick, C.O. Heinke, A. Bodaghee,
S.E. Boggs, P. Kaaret, S. Chaty, J. Rodriguez, and
R. Walter, {\tt arXiv:1202.1531v1 [astro-ph.HE]} (2012).

\bibitem[{\citenamefont{Gudmundsson
et~al.}(1983)\citenamefont{Gudmundsson,
  Pethick, and Epstein}}]{Gudmundsson1983}
\bibinfo{author}{\bibfnamefont{E.~H.} \bibnamefont{Gudmundsson}},
  \bibinfo{author}{\bibfnamefont{C.~J.} \bibnamefont{Pethick}},
  \bibnamefont{and} \bibinfo{author}{\bibfnamefont{R.~I.}
  \bibnamefont{Epstein}}, \bibinfo{journal}{The Astrophysical Journal}
  \textbf{\bibinfo{volume}{272}}, \bibinfo{pages}{286}
(\bibinfo{year}{1983}).

\bibitem[{\citenamefont{Potekhin et~al.}(1997)\citenamefont{Potekhin,
Chabrier,
  and Yakovlev}}]{Potekhin1997}
\bibinfo{author}{\bibfnamefont{A.~Y.} \bibnamefont{Potekhin}},
  \bibinfo{author}{\bibfnamefont{G.}~\bibnamefont{Chabrier}},
\bibnamefont{and}
  \bibinfo{author}{\bibfnamefont{D.~G.} \bibnamefont{Yakovlev}},
  \bibinfo{journal}{Astronomy and Astrophysics}
\textbf{\bibinfo{volume}{323}},
  \bibinfo{pages}{415} (\bibinfo{year}{1997}).

\end{thebibliography}

\end{document}